\documentclass[final,5p,times,twocolumn,authoryear]{elsarticle}
\usepackage{makecell}
\usepackage{graphicx}
\usepackage{caption}
\usepackage{subcaption}
\usepackage{multirow} 
\usepackage{booktabs}
\usepackage[table]{xcolor}
\usepackage{colortbl}
\usepackage{amssymb}
\usepackage{geometry}
\usepackage{amssymb}
\usepackage{lipsum}
\usepackage{xcolor}

\usepackage{booktabs}

\makeatletter
\def\ps@pprintTitle{%
   \let\@oddhead\@empty
   \let\@evenhead\@empty
   \let\@oddfoot\@empty
   \let\@evenfoot\@empty
}
\makeatother

\begin{document}

\begin{frontmatter}

\title{Prostate Cancer Screening with Artificial Intelligence–Enhanced Micro-Ultrasound: A Comparative Study with Traditional Methods}

\author[1]{Muhammad Imran}
\author[2]{Wayne G. Brisbane}
\author[3]{Li-Ming Su}
\author[3]{Jason P. Joseph}
\author[1]{Wei Shao\corref{cor1}}
\cortext[cor1]{Corresponding author. \\
\hspace*{1.9em}E-mail address: weishao@ufl.edu (W. Shao) \\
\hspace*{1.9em}ORCID: 0000-0003-4931-4839 (W. Shao)}

\affiliation[1]{organization={Department of Medicine, University of Florida},
            addressline={}, 
            city={Gainesville},
            postcode={32611}, 
            state={FL},
            country={USA}}
\affiliation[2]{organization={Department of  Urology, University of California},
            addressline={}, 
            city={Los Angeles},
            postcode={90095}, 
            state={CA},
            country={USA}}
\affiliation[3]{organization={Department of Urology, University of Florida},
            addressline={}, 
            city={Gainesville},
            postcode={32611}, 
            state={FL},
            country={USA}}

\begin{abstract}
\textcolor{orange}{\textbf{\textit{Background and objective:}}}
Micro-ultrasound (micro-US) is a novel ultrasound modality with diagnostic accuracy comparable to magnetic resonance imaging (MRI) for detecting clinically significant prostate cancer (csPCa). This study investigated whether interpretation of micro-US by artificial intelligence (AI) can outperform clinical screening methods using prostate-specific antigen (PSA) and digital rectal examination (DRE).
\\
\textcolor{orange}{\textbf{\textit{Methods:}}} 
We retrospectively studied 145 men who underwent micro-US guided biopsy (79 with csPCa and 66 without). A self-supervised convolutional autoencoder was trained to extract deep image features from 2D micro-US slices. Random forest classifiers were developed using five-fold cross-validation to predict csPCa at the slice level. A patient was classified as csPCa-positive if $\ge$8 consecutive slices were predicted positive. Model performance was compared with a classifier trained on common clinical screening variables (PSA, DRE, prostate volume, and age). \\
\textcolor{orange}{\textbf{\textit{Key findings and limitations:}}} 
The AI-enhanced micro-US model and the clinical screening model achieved AUROCs of 0.871 and 0.753, respectively. Using a fixed classification threshold for both models, the micro-US model achieved a sensitivity of 92.5\% and specificity of 68.1\%, while the clinical model achieved a sensitivity of 96.2\% but with a lower specificity at 27.3\%. Limitations of this study include its retrospective single-center design and lack of external validation. \\
\textcolor{orange}{\textbf{\textit{Conclusions and clinical implications:}}} 
AI-interpreted micro-US significantly improves specificity for csPCa while maintaining high sensitivity. This approach may reduce unnecessary biopsies and offers a low-cost, point-of-care alternative to PSA-based screening. Future prospective studies are needed to validate these findings. \\
\textcolor{orange}{\textbf{\textit{Patient summary:}}} We developed an artificial intelligence system to analyze micro-ultrasound images of the prostate. In this study, it detected aggressive prostate cancer more accurately than traditional screening methods such as PSA blood tests and digital rectal exams. This approach may help reduce unnecessary biopsies in the future.
\end{abstract}


\end{frontmatter}

\section{Introduction}
\label{introduction}
Prostate cancer is one of the most commonly diagnosed malignancies and a leading cause of cancer-related death worldwide~\citep{james2024lancet}. 
Early detection of clinically significant prostate cancer (csPCa) is critical, as it increases the 5-year survival rate from 37\% to nearly 100\%~\citep{american2025cancer}. Clinically, screening often relies on prostate-specific antigen (PSA) testing and digital rectal examination (DRE), but both methods have notable limitations. PSA lacks specificity and may be elevated in benign conditions such as benign prostatic hyperplasia or prostatitis. DRE, while inexpensive and easy to perform, suffers from poor sensitivity and high inter-observer variability. As a result, these traditional tools can lead to both overdiagnosis and missed diagnoses, contributing to unnecessary biopsies or delayed detection of aggressive disease.

Multiparametric magnetic resonance imaging (mpMRI) improves the detection of csPCa and is commonly used to guide targeted biopsies~\citep{ahmed2017diagnostic}. However, its role in routine screening is limited by high cost, long acquisition times, limited availability, and the need for specialized radiological expertise. These constraints make mpMRI impractical for large-scale or point-of-care screening. Micro-ultrasound (micro-US), a high-resolution (29 MHz) imaging modality, provides real-time visualization of prostate microarchitecture with spatial resolution three to four times greater than conventional transrectal ultrasound~\citep{klotz2020comparison}. The OPTIMUM randomized trial confirmed that micro-US–guided biopsy is noninferior to MRI-targeted biopsy for detecting csPCa~\citep{kinnaird2025microultrasonography}. With its portability, lower cost, and suitability for outpatient use, micro-US is well positioned as a potential screening tool. However, interpretation remains challenging and highly operator-dependent~\citep{zhou2024inter}, limiting consistent performance and widespread adoption.

To address the interpretive limitations of micro-ultrasound, we developed an artificial intelligence (AI) model to automatically detect csPCa from micro-US images. Using a retrospective cohort of 145 men who underwent micro-US–guided prostate biopsy, we trained a self-supervised convolutional autoencoder to extract deep imaging features from 2D micro-US slices. These features were used to train a random forest classifier for slice-level prediction, and patient-level prediction was determined by aggregating predictions across consecutive slices. We compared this model to a classifier trained on standard clinical screening variables including PSA, DRE, prostate volume, and age. To our knowledge, this is the first study to align micro-US imaging with biopsy-confirmed pathology at the slice level and to perform patient-level csPCa screening predictions using AI. This work evaluates the potential of AI-enhanced micro-US as a practical and accurate tool for prostate cancer screening.

\section{Patients and methods}
\label{methods}
\subsection{Patient population and data description}
This retrospective study was approved by the University of Florida Institutional Review Board and included 145 men who underwent micro-ultrasound (micro-US)–guided prostate biopsy. All patients had clinical indications for biopsy, such as elevated PSA and/or an abnormal digital rectal examination (DRE). Most patients also underwent a systematic 12-core biopsy, including those without visible micro-US lesions. During biopsy, the operator recorded needle trajectories and target locations using micro-US images, enabling retrospective mapping of cores to corresponding regions. All patients provided informed consent. Ground truth for csPCa was established by histopathological analysis of all biopsy cores. Patients were classified as csPCa-positive if any core contained Gleason score $\ge$3+4. Baseline patient characteristics are summarized in Table~\ref{tab:cohort-details}.

\begin{table*}[htbp]
\centering
\caption{Baseline characteristics of our study cohort. Values are medians with interquartile ranges (IQRs) or counts with percentages.}
\label{tab:cohort-details}
\small 
\definecolor{headercolor}{RGB}{255, 204, 153} 
\definecolor{bottomcolor}{RGB}{204, 255, 204} 
\rowcolors{1}{gray!10}{white}

\begin{tabular}{lcc}
\specialrule{.12em}{.1em}{.1em}
\rowcolor{headercolor}
\textbf{Characteristic} & \textbf{Positive Cases} & \textbf{Negative Cases} \\
\specialrule{.12em}{.1em}{.1em}
Age (yr), median (IQR) & 70 (66–74) & 69 (63–71) \\
PSA (ng/ml), median (IQR) & 8.2 (5.8–13.1) & 5.7 (3.3–7.5) \\
DRE = 1, n (\%) & 39 (49.4\%) & 6 (9.1\%) \\
DRE = 0, n (\%) & 40 (50.6\%) & 60 (90.9\%) \\
Prostate volume (ml), median (IQR) & 37.5 (31.5–49.4) & 47.1 (39.0–55.4) \\
\specialrule{.12em}{.1em}{.1em}
\rowcolor{headercolor}
\textbf{Total} & 79 & 66 \\
\specialrule{.12em}{.1em}{.1em}
\end{tabular}
\end{table*}

\subsubsection{Micro-ultrasound imaging}
Pre-biopsy micro-US scans were acquired using a 29 MHz transrectal system (ExactVu, Exact Imaging, Markham, Canada) by an experienced urologist (WGB) with four years of micro-US interpretation experience. Scans were recorded at 10 frames per second for up to 30 seconds, producing 200 to 300 2D micro-US slices per scan.

\subsubsection{Clinical biomarkers}
For each patient, we collected the following clinical biomarkers: PSA, DRE findings, age, and prostate volume. Prostate volume was estimated from the pre-biopsy micro-US scan. We used the MicroSegNet model~\citep{jiang2024microsegnet} to segment the prostate capsule on each 2D slice, and reconstructed the segmentations into a 3D volume using the method described in~\citep{imran2024image}. The final prostate gland volume (in mL) was computed from the reconstructed 3D model.

\subsubsection{Slice-level labeling for model training}
Model training required slice-level annotations indicating the presence or absence of csPCa on each 2D micro-ultrasound (micro-US) slice. For csPCa-negative patients, all slices were labeled as negative. For csPCa-positive patients, operator-recorded needle trajectories were used to cognitively map each biopsy core to the corresponding region on the pre-biopsy micro-US scan. An expert urologist (WGB) manually reviewed slices surrounding these trajectories to assess csPCa extent, using sonographic features defined in the PRI-MUS protocol~\citep{maffei2024mp49}. Slices with suspicious features were labeled as positive. All other slices in csPCa-positive cases were excluded from training, as their cancer status could not be confidently determined in the absence of histopathological confirmation. In total, 2,062 positive and 14,769 negative slices were included. Model evaluation was conducted at the patient level, using biopsy-confirmed csPCa status to reflect clinically meaningful outcomes.
\subsection{Model development and evaluation}
\subsubsection{Micro-US image feature extraction}
\label{sect:feature-learning-via-convolutional-autoencoder}
\begin{figure}[ht]
    \centering
        \includegraphics[width=0.45\textwidth]{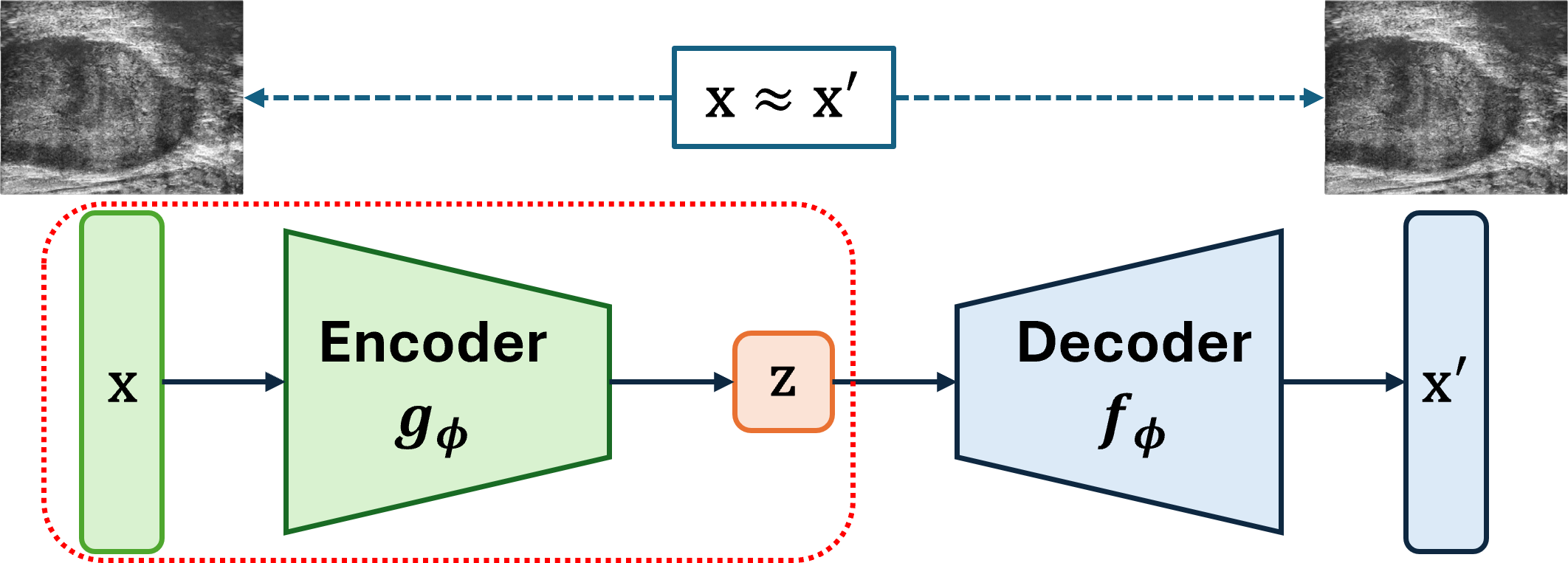}
        \caption{Architecture of the convolutional autoencoder used for feature extraction.}
        \label{fig:feature-learning-via-convolutional-autoencoder}
\end{figure}

We trained a convolutional autoencoder with self-supervised learning to extract high-level features from micro-US images.
The autoencoder (Figure~\ref{fig:feature-learning-via-convolutional-autoencoder}) consists of two symmetric components: a convolutional encoder $g_\phi$ and a decoder $f_\phi$. The encoder compresses the input image $\mathbf{x}$ into a lower-dimensional latent representation $\mathbf{z}$, while the decoder attempts to reconstruct the original image from $\mathbf{z}$. The encoder includes five convolutional layers with increasing channel dimensions (from 3 to 256), interleaved with ReLU activations and strided convolutions for spatial downsampling. The decoder mirrors this structure with transposed convolutions and corresponding upsampling layers to produce the reconstructed image. The autoencoder was trained to minimize the mean squared error between the input image $\mathbf{x}$ and the reconstructed output $\mathbf{x'}$. 
After training, we used the encoder as a fixed feature extractor. Each 2D micro-US slice was passed through the encoder to generate a feature map, which was then reduced via adaptive average pooling to obtain a 256-dimensional feature vector.

\subsubsection{Micro-US image classification}
We trained random forest classifiers to classify 2D micro-US slices as csPCa-positive or negative using 256-dimensional feature vectors extracted by the autoencoder’s encoder, which captured key imaging characteristics such as texture, intensity, and shape. Slice-level predictions were aggregated to produce patient-level classifications. A patient was considered csPCa-positive if at least eight consecutive slices were predicted positive. This rule was based on retrospective analysis of lesion length, which showed that csPCa typically spanned eight adjacent slices on average. Requiring spatially contiguous predictions helped reduce false positives and improved specificity without compromising sensitivity.

\subsubsection{Classification with clinical biomarkers}
To assess the predictive value of commonly used screening tools, we trained a random forest model using only non-imaging features: patient age, PSA level, prostate volume, and the binary outcome of DRE.
Since the \texttt{scikit-learn} implementation supports internal out-of-bag (OOB) validation, a separate validation set was not required for hyperparameter tuning.

\subsubsection{Cross-validation strategy}
We used five-fold cross-validation to ensure robust and unbiased model evaluation. The dataset, consisting of 79 csPCa-positive and 66 csPCa-negative patients, was partitioned into five mutually exclusive folds. Each patient appeared exactly once in the test set, once in the validation set, and in the training set for the remaining three folds. This approach ensured that all cases contributed to both training and evaluation while preventing data leakage. For each fold, training was monitored on the validation set to prevent overfitting, and the model checkpoint with the lowest validation loss was retained.

\subsubsection{Implementation details}
All autoencoder models were implemented in \texttt{PyTorch (v1.13)} and trained on an NVIDIA A100 GPU using the Adam optimizer (learning rate = 0.001, batch size = 32). After training, the decoder was discarded and the encoder was used as a fixed feature extractor. 
Each random forest model was trained with 1,000 trees, class-balanced weights, and stratified sampling to preserve the distribution of positive and negative slices. Records with missing clinical values or duplicate patient entries were removed to ensure data integrity.

\subsubsection{Peformance metrics}
We evaluated model performance at the patient level using the following metrics: area under the receiver operating characteristic curve (AUROC), accuracy, sensitivity, specificity, precision, and F1-score. All metrics were averaged across the five cross-validation folds. AUROC served as the primary evaluation metric, as it captures overall discriminatory performance across all classification thresholds. The remaining metrics were computed using a fixed probability threshold of 0.15, which was empirically selected to balance sensitivity and specificity in the training data.

\section{Results}

\subsection{ROC-based comparison of imaging and clinical models}

\begin{figure}[ht]
    \centering
    \includegraphics[width=0.48\textwidth]{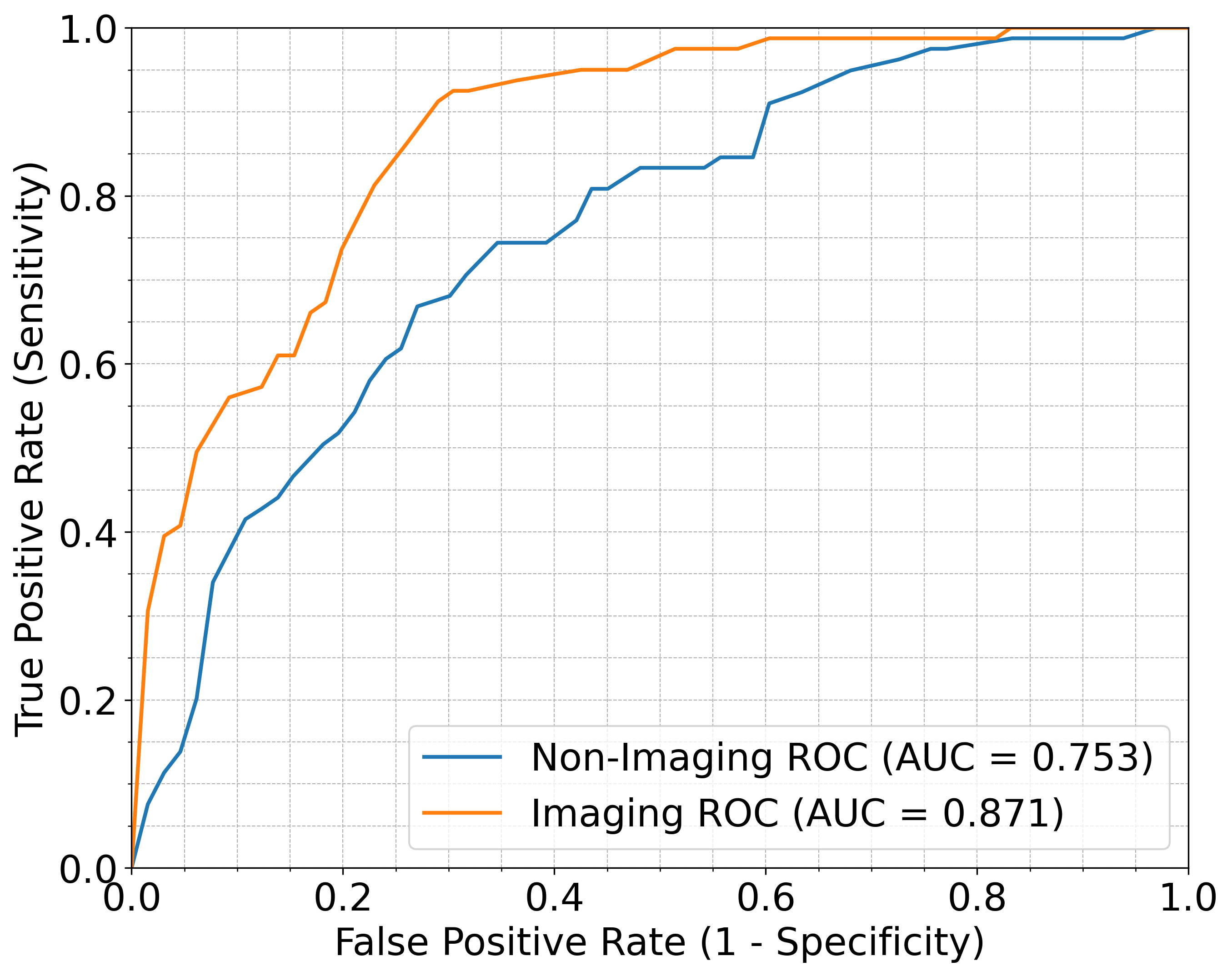}
    \caption{ROC curves comparing the imaging-based model and clinical biomarker-based model. The imaging model achieved a higher AUROC (0.871) than the clinical model (0.753).}
    \label{fig:roc-curves}
\end{figure}

Figure~\ref{fig:roc-curves} shows the ROC curves comparing the performance of the imaging-based model and the clinical biomarker model. The clinical model, trained on age, PSA, DRE, and prostate volume, achieved a mean AUROC of 0.753, indicating moderate discriminative ability. In contrast, the imaging-based model achieved a higher AUROC of 0.871, reflecting a stronger ability to distinguish csPCa from non-csPCa based on deep features extracted from micro-US slices.

\subsection{Comparison of classification metrics}

Table~\ref{tab:results-comparison} summarizes the average classification metrics across five cross-validation folds. The clinical model achieved high sensitivity (96.2\%) but low specificity (27.3\%), resulting in a high false-positive rate. Its precision (61.4\%), F1-score (74.9\%), and accuracy (64.8\%) were moderate. In contrast, the imaging model maintained high sensitivity (92.5\%) while substantially improving specificity (68.1\%). It also achieved higher precision (77.8\%), F1-score (84.5\%), and accuracy (81.4\%), demonstrating a better overall balance between true-positive detection and false-positive control.

\begin{table}[ht]
    \centering
    \caption{Threshold-based classification metrics (averaged over five folds) using a fixed decision threshold of 0.15.}
    \definecolor{headercolor}{RGB}{255, 204, 153} 
    \label{tab:results-comparison}
    \resizebox{\columnwidth}{!}{%
        \begin{tabular}{lcccccc}
            \specialrule{.12em}{.1em}{.1em} 
            \rowcolor{headercolor}
            \textbf{Model} & \textbf{Sensitivity} & \textbf{Specificity} & \textbf{Accuracy} & \textbf{Precision} & \textbf{F1-Score} \\
            \specialrule{.12em}{.1em}{.1em} 
            Clinical  & \textbf{96.2\%} & 27.3\% & 64.8\% & 61.4\% & 74.9\% \\
            Micro-US & 92.5\% & \textbf{68.1\%} & \textbf{81.4\%} & \textbf{77.8\%} & \textbf{84.5\%} \\
            \specialrule{.12em}{.1em}{.1em} 
        \end{tabular}
    }
\end{table}

These results highlight that the AI-based imaging model offers a more favorable trade-off between sensitivity and specificity, reducing unnecessary false positives while maintaining robust detection of clinically significant disease. This supports its potential as a more accurate and efficient screening tool for csPCa.

\section{Discussion}

This study demonstrates that an AI-enhanced micro-US model can significantly improve prostate cancer screening performance compared to traditional biomarker-based approaches. While the clinical model, which incorporated PSA, DRE, prostate volume, and age, achieved high sensitivity (96.2\%), it exhibited poor specificity (27.3\%), consistent with well-documented limitations of PSA-based screening~\citep{thompson2004prevalence, schroder1998evaluation}. In contrast, the imaging-based model maintained high sensitivity (92.5\%) while substantially improving specificity (68.1\%), resulting in better overall accuracy and precision. This improved balance is particularly important in a screening context, where reducing false positives can lower the burden of unnecessary biopsies, overtreatment, and patient anxiety.

Micro-US has been proposed as a lower-cost, point-of-care alternative to multiparametric MRI for prostate cancer detection~\citep{klotz2020comparison, lughezzani2019comparison}. Prior studies, including the OPTIMUM randomized trial, have shown that micro-US is non-inferior to MRI for csPCa detection in biopsy settings~\citep{kinnaird2025microultrasonography}. However, its broader use in screening has been limited by high inter-operator variability and a steep learning curve~\citep{zhou2024inter}. Our results suggest that artificial intelligence can help overcome these barriers by enabling consistent, objective interpretation of micro-US images. The self-supervised autoencoder used in this study learned imaging features correlated with csPCa, which were then aggregated using a slice-level prediction framework to produce patient-level classifications. This approach effectively eliminates reliance on subjective interpretation and may help standardize micro-US for widespread clinical adoption.

Previous models, such as TRUSformer~\citep{gilany2023trusformer} and TRUSWorthy~\citep{harmanani2025trusworthy}, have focused on patch-level classification using weak labels, limiting their clinical interpretability. In contrast, our model aligned individual micro-US slices with biopsy-confirmed pathology and used an empirically chosen aggregation rule to generate patient-level predictions. This method mirrors clinical decision-making and supports actionable, real-time screening decisions. The improved diagnostic performance, particularly in specificity and F1-score, underscores the potential of AI-enhanced micro-US to function as a frontline screening modality that complements or even outperforms existing PSA- and DRE-based strategies.

Despite encouraging results, this study has limitations. The cohort was drawn from a single academic center and consisted of patients already referred for biopsy, which may introduce selection bias. As such, performance in general screening populations remains to be validated. While five-fold cross-validation was used to minimize overfitting, external validation on independent cohorts is necessary to assess generalizability. Additionally, the threshold for patient-level classification (eight consecutive positive slices) was empirically defined and may require adjustment in future prospective settings. Future work should include multicenter clinical trials, decision-curve analysis, and cost-effectiveness studies to evaluate the real-world impact of integrating AI-enhanced micro-US into routine prostate cancer screening. If validated, this approach could provide a scalable, affordable, and interpretable solution for early detection of clinically significant prostate cancer, bridging the gap between low-specificity biomarker screening and high-cost MRI-based diagnostics.

\section{Conclusions}
\label{conclusion}
This study indicates that AI-augmented micro-ultrasound can outperform traditional PSA- and DRE-based methods for screening clinically significant prostate cancer. These results highlight the potential of micro-US, when interpreted by AI, to serve as a point-of-care screening tool that detects clinically significant cancers more accurately and reduces unnecessary biopsies. If validated in prospective multi-center settings, AI-enhanced micro-US could transform early prostate cancer detection by enabling more precise, accessible, and cost-effective screening, ultimately improving patient outcomes while minimizing harm.
\newline

\noindent\textbf{\textit{Financial disclosures:}} Wei Shao certifies that all conflicts of interest, including specific financial interests and relationships and affiliations relevant to the subject matter or materials discussed in the manuscript (e.g., employment/affiliation, grants or funding, consultancies, honoraria, stock ownership or options, expert testimony, royalties, or patents filed,
received, or pending), are the following: None.\\

\noindent\textbf{\textit{Funding/Support and role of the sponsor:}} This work was supported by the Department of Medicine and the Intelligent Clinical Care Center at the University of Florida College of Medicine. The authors express their sincere gratitude to the NVIDIA AI Technology Center at the University of Florida for their invaluable feedback, technical guidance, and support throughout this project.

\bibliographystyle{elsarticle-harv} 
\bibliography{example}

\begin{thebibliography}{14}
\expandafter\ifx\csname natexlab\endcsname\relax\def\natexlab#1{#1}\fi
\providecommand{\url}[1]{\texttt{#1}}
\providecommand{\href}[2]{#2}
\providecommand{\path}[1]{#1}
\providecommand{\DOIprefix}{doi:}
\providecommand{\ArXivprefix}{arXiv:}
\providecommand{\URLprefix}{URL: }
\providecommand{\Pubmedprefix}{pmid:}
\providecommand{\doi}[1]{\href{http://dx.doi.org/#1}{\path{#1}}}
\providecommand{\Pubmed}[1]{\href{pmid:#1}{\path{#1}}}
\providecommand{\bibinfo}[2]{#2}
\ifx\xfnm\relax \def\xfnm[#1]{\unskip,\space#1}\fi
\bibitem[{Ahmed et~al.(2017)Ahmed, Bosaily, Brown, Gabe, Kaplan, Parmar, Collaco-Moraes, Ward, Hindley, Freeman et~al.}]{ahmed2017diagnostic}
\bibinfo{author}{Ahmed, H.U.}, \bibinfo{author}{Bosaily, A.E.S.}, \bibinfo{author}{Brown, L.C.}, \bibinfo{author}{Gabe, R.}, \bibinfo{author}{Kaplan, R.}, \bibinfo{author}{Parmar, M.K.}, \bibinfo{author}{Collaco-Moraes, Y.}, \bibinfo{author}{Ward, K.}, \bibinfo{author}{Hindley, R.G.}, \bibinfo{author}{Freeman, A.}, et~al., \bibinfo{year}{2017}.
\newblock \bibinfo{title}{Diagnostic accuracy of multi-parametric mri and trus biopsy in prostate cancer (promis): a paired validating confirmatory study}.
\newblock \bibinfo{journal}{The Lancet} \bibinfo{volume}{389}, \bibinfo{pages}{815--822}.
\bibitem[{Gilany et~al.(2023)Gilany, Wilson, Perera-Ortega, Jamzad, To, Fooladgar, Wodlinger, Abolmaesumi and Mousavi}]{gilany2023trusformer}
\bibinfo{author}{Gilany, M.}, \bibinfo{author}{Wilson, P.}, \bibinfo{author}{Perera-Ortega, A.}, \bibinfo{author}{Jamzad, A.}, \bibinfo{author}{To, M.N.N.}, \bibinfo{author}{Fooladgar, F.}, \bibinfo{author}{Wodlinger, B.}, \bibinfo{author}{Abolmaesumi, P.}, \bibinfo{author}{Mousavi, P.}, \bibinfo{year}{2023}.
\newblock \bibinfo{title}{Trusformer: improving prostate cancer detection from micro-ultrasound using attention and self-supervision}.
\newblock \bibinfo{journal}{International Journal of Computer Assisted Radiology and Surgery} \bibinfo{volume}{18}, \bibinfo{pages}{1193--1200}.
\bibitem[{Harmanani et~al.(2025)Harmanani, Wilson, To, Gilany, Jamzad, Fooladgar, Wodlinger, Abolmaesumi and Mousavi}]{harmanani2025trusworthy}
\bibinfo{author}{Harmanani, M.}, \bibinfo{author}{Wilson, P.F.}, \bibinfo{author}{To, M.N.N.}, \bibinfo{author}{Gilany, M.}, \bibinfo{author}{Jamzad, A.}, \bibinfo{author}{Fooladgar, F.}, \bibinfo{author}{Wodlinger, B.}, \bibinfo{author}{Abolmaesumi, P.}, \bibinfo{author}{Mousavi, P.}, \bibinfo{year}{2025}.
\newblock \bibinfo{title}{Trusworthy: toward clinically applicable deep learning for confident detection of prostate cancer in micro-ultrasound}.
\newblock \bibinfo{journal}{International Journal of Computer Assisted Radiology and Surgery} , \bibinfo{pages}{1--9}.
\bibitem[{Imran et~al.(2024)Imran, Nguyen, Pensa, Falzarano, Sisk, Liang, DiBianco, Su, Zhou, Joseph et~al.}]{imran2024image}
\bibinfo{author}{Imran, M.}, \bibinfo{author}{Nguyen, B.}, \bibinfo{author}{Pensa, J.}, \bibinfo{author}{Falzarano, S.M.}, \bibinfo{author}{Sisk, A.E.}, \bibinfo{author}{Liang, M.}, \bibinfo{author}{DiBianco, J.M.}, \bibinfo{author}{Su, L.M.}, \bibinfo{author}{Zhou, Y.}, \bibinfo{author}{Joseph, J.P.}, et~al., \bibinfo{year}{2024}.
\newblock \bibinfo{title}{Image registration of in vivo micro-ultrasound and ex vivo pseudo-whole mount histopathology images of the prostate: A proof-of-concept study}.
\newblock \bibinfo{journal}{Biomedical Signal Processing and Control} \bibinfo{volume}{96}, \bibinfo{pages}{106657}.
\bibitem[{James et~al.(2024)James, Tannock, N'Dow, Feng, Gillessen, Ali, Trujillo, Al-Lazikani, Attard, Bray et~al.}]{james2024lancet}
\bibinfo{author}{James, N.D.}, \bibinfo{author}{Tannock, I.}, \bibinfo{author}{N'Dow, J.}, \bibinfo{author}{Feng, F.}, \bibinfo{author}{Gillessen, S.}, \bibinfo{author}{Ali, S.A.}, \bibinfo{author}{Trujillo, B.}, \bibinfo{author}{Al-Lazikani, B.}, \bibinfo{author}{Attard, G.}, \bibinfo{author}{Bray, F.}, et~al., \bibinfo{year}{2024}.
\newblock \bibinfo{title}{The lancet commission on prostate cancer: planning for the surge in cases}.
\newblock \bibinfo{journal}{The Lancet} \bibinfo{volume}{403}, \bibinfo{pages}{1683--1722}.
\bibitem[{Jiang et~al.(2024)Jiang, Imran, Muralidharan, Patel, Pensa, Liang, Benidir, Grajo, Joseph, Terry et~al.}]{jiang2024microsegnet}
\bibinfo{author}{Jiang, H.}, \bibinfo{author}{Imran, M.}, \bibinfo{author}{Muralidharan, P.}, \bibinfo{author}{Patel, A.}, \bibinfo{author}{Pensa, J.}, \bibinfo{author}{Liang, M.}, \bibinfo{author}{Benidir, T.}, \bibinfo{author}{Grajo, J.R.}, \bibinfo{author}{Joseph, J.P.}, \bibinfo{author}{Terry, R.}, et~al., \bibinfo{year}{2024}.
\newblock \bibinfo{title}{Microsegnet: A deep learning approach for prostate segmentation on micro-ultrasound images}.
\newblock \bibinfo{journal}{Computerized Medical Imaging and Graphics} \bibinfo{volume}{112}, \bibinfo{pages}{102326}.
\bibitem[{Kinnaird et~al.(2025)Kinnaird, Luger, Cash, Ghai, Urdaneta-Salegui, Pavlovich, Brito, Shore, Struck, Schostak et~al.}]{kinnaird2025microultrasonography}
\bibinfo{author}{Kinnaird, A.}, \bibinfo{author}{Luger, F.}, \bibinfo{author}{Cash, H.}, \bibinfo{author}{Ghai, S.}, \bibinfo{author}{Urdaneta-Salegui, L.F.}, \bibinfo{author}{Pavlovich, C.P.}, \bibinfo{author}{Brito, J.}, \bibinfo{author}{Shore, N.D.}, \bibinfo{author}{Struck, J.P.}, \bibinfo{author}{Schostak, M.}, et~al., \bibinfo{year}{2025}.
\newblock \bibinfo{title}{Microultrasonography-guided vs mri-guided biopsy for prostate cancer diagnosis: The optimum randomized clinical trial}.
\newblock \bibinfo{journal}{JAMA} .
\bibitem[{Klotz et~al.(2020)Klotz, Lughezzani, Maffei, S{\'a}nchez, Pereira, Staerman, Cash, Luger, Lopez, Sanchez-Salas et~al.}]{klotz2020comparison}
\bibinfo{author}{Klotz, L.}, \bibinfo{author}{Lughezzani, G.}, \bibinfo{author}{Maffei, D.}, \bibinfo{author}{S{\'a}nchez, A.}, \bibinfo{author}{Pereira, J.G.}, \bibinfo{author}{Staerman, F.}, \bibinfo{author}{Cash, H.}, \bibinfo{author}{Luger, F.}, \bibinfo{author}{Lopez, L.}, \bibinfo{author}{Sanchez-Salas, R.}, et~al., \bibinfo{year}{2020}.
\newblock \bibinfo{title}{Comparison of micro-ultrasound and multiparametric magnetic resonance imaging for prostate cancer: A multicenter, prospective analysis}.
\newblock \bibinfo{journal}{Canadian Urological Association Journal} \bibinfo{volume}{15}, \bibinfo{pages}{E11}.
\bibitem[{Lughezzani et~al.(2019)Lughezzani, Saita, Lazzeri, Paciotti, Maffei, Lista, Hurle, Buffi, Guazzoni and Casale}]{lughezzani2019comparison}
\bibinfo{author}{Lughezzani, G.}, \bibinfo{author}{Saita, A.}, \bibinfo{author}{Lazzeri, M.}, \bibinfo{author}{Paciotti, M.}, \bibinfo{author}{Maffei, D.}, \bibinfo{author}{Lista, G.}, \bibinfo{author}{Hurle, R.}, \bibinfo{author}{Buffi, N.M.}, \bibinfo{author}{Guazzoni, G.}, \bibinfo{author}{Casale, P.}, \bibinfo{year}{2019}.
\newblock \bibinfo{title}{Comparison of the diagnostic accuracy of micro-ultrasound and magnetic resonance imaging/ultrasound fusion targeted biopsies for the diagnosis of clinically significant prostate cancer}.
\newblock \bibinfo{journal}{European urology oncology} \bibinfo{volume}{2}, \bibinfo{pages}{329--332}.
\bibitem[{Maffei et~al.(2024)Maffei, Avolio, Moretto, Piccolini, Aljoulani, Dagnino, De~Carne, Fasulo, Marco, Saita et~al.}]{maffei2024mp49}
\bibinfo{author}{Maffei, D.}, \bibinfo{author}{Avolio, P.P.}, \bibinfo{author}{Moretto, S.}, \bibinfo{author}{Piccolini, A.}, \bibinfo{author}{Aljoulani, M.}, \bibinfo{author}{Dagnino, F.}, \bibinfo{author}{De~Carne, F.}, \bibinfo{author}{Fasulo, V.}, \bibinfo{author}{Marco, P.}, \bibinfo{author}{Saita, A.R.}, et~al., \bibinfo{year}{2024}.
\newblock \bibinfo{title}{Mp49-15 evaluating the role of pri-mus protocol in identifying clinically significant prostate cancer: A high-volume experience on microultrasound}.
\newblock \bibinfo{journal}{Journal of Urology} \bibinfo{volume}{211}, \bibinfo{pages}{e788}.
\bibitem[{Schr{\"o}der et~al.(1998)Schr{\"o}der, Kruger, Rietbergen, Kranse, Maas, Beemsterboer and Hoedemaeker}]{schroder1998evaluation}
\bibinfo{author}{Schr{\"o}der, F.H.}, \bibinfo{author}{Kruger, A.B.}, \bibinfo{author}{Rietbergen, J.}, \bibinfo{author}{Kranse, R.}, \bibinfo{author}{Maas, P.v.d.}, \bibinfo{author}{Beemsterboer, P.}, \bibinfo{author}{Hoedemaeker, R.}, \bibinfo{year}{1998}.
\newblock \bibinfo{title}{Evaluation of the digital rectal examination as a screening test for prostate cancer}.
\newblock \bibinfo{journal}{Journal of the National Cancer Institute} \bibinfo{volume}{90}, \bibinfo{pages}{1817--1823}.
\bibitem[{Society(2025)}]{american2025cancer}
\bibinfo{author}{Society, A.C.}, \bibinfo{year}{2025}.
\newblock \bibinfo{title}{Cancer facts \& figures 2025. atlanta: American cancer society; 2025.}
\bibitem[{Thompson et~al.(2004)Thompson, Pauler, Goodman, Tangen, Lucia, Parnes, Minasian, Ford, Lippman, Crawford et~al.}]{thompson2004prevalence}
\bibinfo{author}{Thompson, I.M.}, \bibinfo{author}{Pauler, D.K.}, \bibinfo{author}{Goodman, P.J.}, \bibinfo{author}{Tangen, C.M.}, \bibinfo{author}{Lucia, M.S.}, \bibinfo{author}{Parnes, H.L.}, \bibinfo{author}{Minasian, L.M.}, \bibinfo{author}{Ford, L.G.}, \bibinfo{author}{Lippman, S.M.}, \bibinfo{author}{Crawford, E.D.}, et~al., \bibinfo{year}{2004}.
\newblock \bibinfo{title}{Prevalence of prostate cancer among men with a prostate-specific antigen level $\le$ 4.0 ng per milliliter}.
\newblock \bibinfo{journal}{New England Journal of Medicine} \bibinfo{volume}{350}, \bibinfo{pages}{2239--2246}.
\bibitem[{Zhou et~al.(2024)Zhou, Choi, Vesal, Kinnaird, Brisbane, Lughezzani, Maffei, Fasulo, Albers, Zhang et~al.}]{zhou2024inter}
\bibinfo{author}{Zhou, S.R.}, \bibinfo{author}{Choi, M.H.}, \bibinfo{author}{Vesal, S.}, \bibinfo{author}{Kinnaird, A.}, \bibinfo{author}{Brisbane, W.G.}, \bibinfo{author}{Lughezzani, G.}, \bibinfo{author}{Maffei, D.}, \bibinfo{author}{Fasulo, V.}, \bibinfo{author}{Albers, P.}, \bibinfo{author}{Zhang, L.}, et~al., \bibinfo{year}{2024}.
\newblock \bibinfo{title}{Inter-reader agreement for prostate cancer detection using micro-ultrasound: a multi-institutional study}.
\newblock \bibinfo{journal}{European Urology Open Science} \bibinfo{volume}{66}, \bibinfo{pages}{93--100}.

\end{thebibliography}
\end{document}